\documentstyle[12pt]{article}
\input epsf
\begin{document}
\rightline{\hbox{EFI-96-29}}
\vskip 1cm
\centerline{{\bf \Large \bf
Geometrical Structures of M-Theory}}

\vskip 1cm
\centerline{{\bf \Large Emil J. Martinec}}
\vskip .4cm
\centerline{\it Enrico Fermi Institute and Dept. of Physics}
\centerline{\it University of Chicago,
5640 S. Ellis Ave., Chicago, IL 60637 USA}


\def\pref#1{(\ref{#1})}

\def\ie{{i.e.}}
\def\eg{{e.g.}}
\def\cf{{c.f.}}
\def\etal{{et.al.}}
\def\etc{{etc.}}

\def\inbar{\,\vrule height1.5ex width.4pt depth0pt}
\def\IR{\relax{\rm I\kern-.18em R}}
\def\IC{\relax\hbox{$\inbar\kern-.3em{\rm C}$}}
\def\IH{\relax{\rm I\kern-.18em H}}
\def\IO{\relax\hbox{$\inbar\kern-.3em{\rm O}$}}
\def\IK{\relax{\rm I\kern-.18em K}}
\def\IP{\relax{\rm I\kern-.18em P}}
\def\Z{{\bf Z}}
\def\Pone{{\IC\rm P^1}}
\def\One{{1\hskip -3pt {\rm l}}}

\def\beq{\begin{equation}}
\def\eeq{\end{equation}}

\def\sst{\scriptscriptstyle}
\def\tst#1{{\textstyle #1}}
\def\frac#1#2{{#1\over#2}}
\def\coeff#1#2{{\textstyle{#1\over #2}}}
\def\half{\frac12}
\def\hf{{\textstyle\half}}
\def\ket#1{|#1\rangle}
\def\bra#1{\langle#1|}
\def\vev#1{\langle#1\rangle}
\def\d{\partial}

\def\tpt{{2+2}}
\def\tentwo{{10+2}}
\def\ij{{i\bar j}}
\def\longhookrightarrow{\lhook\joinrel\longrightarrow}

\def\np{{\it Nucl. Phys. }}
\def\pl{{\it Phys. Lett. }}
\def\pr{{\it Phys. Rev. }}
\def\ap{{\it Ann. Phys., NY }}
\def\prl{{\it Phys. Rev. Lett. }}
\def\mpl{{\it Mod. Phys. Lett. }}
\def\cmp{{\it Comm. Math. Phys. }}
\def\grg{{\it Gen. Rel. and Grav. }}
\def\cqg{{\it Class. Quant. Grav. }}
\def\ijmp{{\it Int. J. Mod. Phys. }}
\def\jmp{{\it J. Math. Phys. }}
\def\nextline{\hfil\break}
\catcode`\@=11
\def\slash#1{\mathord{\mathpalette\c@ncel{#1}}}
\overfullrule=0pt
\def\AA{{\cal A}}
\def\BB{{\cal B}}
\def\CC{{\cal C}}
\def\DD{{\cal D}}
\def\EE{{\cal E}}
\def\FF{{\cal F}}
\def\GG{{\cal G}}
\def\HH{{\cal H}}
\def\II{{\cal I}}
\def\JJ{{\cal J}}
\def\KK{{\cal K}}
\def\LL{{\cal L}}
\def\MM{{\cal M}}
\def\NN{{\cal N}}
\def\OO{{\cal O}}
\def\PP{{\cal P}}
\def\QQ{{\cal Q}}
\def\RR{{\cal R}}
\def\SS{{\cal S}}
\def\TT{{\cal T}}
\def\UU{{\cal U}}
\def\VV{{\cal V}}
\def\WW{{\cal W}}
\def\XX{{\cal X}}
\def\YY{{\cal Y}}
\def\ZZ{{\cal Z}}
\def\lam{\lambda}
\def\eps{\epsilon}
\def\vareps{\varepsilon}
\def\underrel#1\over#2{\mathrel{\mathop{\kern\z@#1}\limits_{#2}}}
\def\lapprox{{\underrel{\scriptstyle<}\over\sim}}
\def\lessapprox{{\buildrel{<}\over{\scriptstyle\sim}}}
\catcode`\@=12

\begin{abstract}
N=(2,1) heterotic string theory provides clues about hidden
structure in M-theory related to string duality; in effect
it geometrizes some aspects of duality.  The program
whereby one may deduce this hidden structure is outlined,
together with the results obtained to date.
Speculations are made as to the eventual shape of the theory.
Talk presented at Strings '96 (Santa Barbara, July 20-25, 1996).
\end{abstract}

\section{\large \bf Introduction}

N=(2,1) heterotic strings seem to know deeply about the duality
structure of M-theory.  In recent work \cite{km,kmo}, it has 
been shown that N=(2,1) strings realize in their target space
all critical string/membrane worldvolume dynamics 
as different background geometries.\footnote{In fact, one
obtains even more than one bargained for -- not only are
IIA/IIB strings, 11D supermembranes, and heterotic/type I strings
obtained \cite{kmo}, but also what appear to be
bosonic strings/membranes \cite{km}, 
the N=(2,2) string, and the N=(2,1) string itself \cite{kmu}!
While I consider this fascinating, I will mostly confine 
myself to spacetime supersymmetric vacua in this talk.}
The overarching structure is \tpt\ dimensional
self-dual geometry, from which the string/membrane dynamics
follows by (null) dimensional reduction.  The conceptual framework
is pictured in figure 1.

\setlength{\unitlength}{1mm}
\begin{picture}(150,60)
\put(0,35){\framebox(30,25){\shortstack{(2,1) string \\ Worldsheet}}}
\put(30,35){\makebox(15,25){\shortstack{\\$\longhookrightarrow$ \\ X(z)}}}
\put(45,35){\framebox(40,25)
	{\shortstack{\tpt\ M-brane \\ Worldvolume \\ (Target Space)}}}
\put(85,35){\makebox(15,25)
	{\shortstack{\\$\longhookrightarrow$ \\ $\Phi$(X)}}}
\put(100,35){\framebox(30,25){\shortstack{\tentwo \\ Spacetime}}}
\put(0,15){\makebox(30,20){\shortstack{Fields X(z)\\
	$S_{\sigma}=\int d^2\hskip -2pt z\;\Phi(X)$}}}
\put(30,15){\makebox(15,20){$\Longrightarrow$ }}
\put(30,0){\makebox(15,20){\shortstack{
	$\beta$-functions, \\ string S-matrix}}}
\put(45,15){\makebox(40,20){\shortstack{Fields $\Phi$(X)\\
	$S_{\tpt}=\int d^4\hskip -2pt X\; \LL(\Phi)$}}}
\put(85,15){\makebox(15,20){$\Longrightarrow$ }}
\put(85,0){\makebox(15,20){\shortstack{
	supersymmetry, \\ $\kappa$-symmetry }}}
\put(100,15){\makebox(30,20){\shortstack{Fields ??\\
	$S_{\tentwo}=??$}}}
\end{picture}

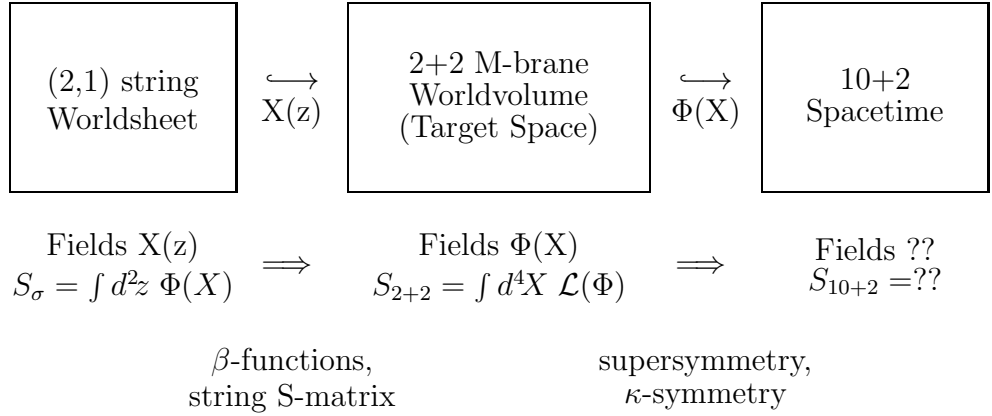
\centerline{Figure 1.  \it The route from N=(2,1) heterotic
strings to M-theory.}
\vskip .3cm

The background fields of the (2,1) string describe the embedding
of a \tpt\ self-dual worldvolume into a \tentwo\ spacetime;
specifically, one finds the \tpt\ worldvolume stretched
over some hyperplane, described in static gauge.
Rather than focussing on the particular properties of any
particular null reduction, we would like to describe the 
general features of these \tpt\ `M-branes',
and what we might learn about the \tentwo\ dimensional
theory they couple to.

There is in fact a concrete procedure to attain this goal,
also indicated in figure 1.  The embedding fields of the
M-brane (or rather, fluctuations in them) are the background
fields of the (2,1) string sigma-model.  Thus the action 
governing the \tpt\ target space dynamics is the one
that generates the N=(2,1) sigma model beta-function equations,
as well as the (2,1) string S-matrix.  This step has recently
been completed in collaboration with D. Kutasov \cite{km2}.
The result is an intriguing mixture of the Dirac-Born-Infeld (DBI)
action for D-branes and the super $p$-brane action:
The exact classical bosonic action is of DBI form:
\beq
  S_{2+2} = \int d^4x\sqrt{\det[\eta_{i\bar j}
	+f_{i\bar j}+\d_i\phi^a\d_{\bar j}\phi^a]}\ .
\label{sfour}
\eeq
A preliminary analysis of the target space fermion couplings
indicates a supersymmetric completion of the sort familiar
from super $p$-branes:
\begin{eqnarray}
  \d_i\phi^a\longrightarrow \d_i\phi^a+\bar\psi\gamma^a\d_i\psi
	\equiv\Pi_i^a\nonumber\\
  f_{i\bar j}\longrightarrow f_{i\bar j}+\bar\psi\gamma_{(i}
	\d_{\bar j)}\psi\equiv \Pi_{i\bar j}\ ,
\label{tanvec}
\end{eqnarray}
together with some sort of Wess-Zumino terms in $S_{2+2}$.
We will describe the derivation of these results in section 2.

Recall that the $p$-brane action has both global spacetime supersymmetry
and $\kappa$-symmetry.  If one did not know what supergravity theory
couples to the eleven-dimensional supermembrane, for example,
one could deduce it by asking for the most general possible
supermembrane action compatible with $\kappa$-symmetry.
This then is the program: find the $\kappa$-symmetric, supersymmetric,
reparametrization invariant action that reduces to $S_{2+2}$
in static gauge; then look for deformations of the
theory compatible with this structure.

The organization of the talk is as follows:
In section 2, we sketch the derivation of the target space 
action \pref{sfour}.  
Section 3 discusses the symmetry properties of the \tpt\
worldvolume theory: its supersymmetry, $\kappa$-symmetry,
self-duality, current algebra, etc.
Section 4 contains some brief remarks about the quantization
of the \tpt\ theory; here analogies with matrix models
of noncritical string theory might prove helpful.  
Section 5 returns to the issue of supersymmetry, and how the
(2,1) string might point towards `$p$-brane democracy'
at short distances.  
Section 6 is an attempt to discern the ultimate structure
of M-theory by extrapolating all these geometrical structures
to their logical conclusion.

\section{\large\bf The \tpt\ target space action}

There are two routes to the target space action of a
string theory, each giving complementary information.
First, the action which generates the sigma-model beta-functions
contains, at a given loop order, all orders in fields
with a fixed number of derivatives.  Second, the generating
function of the string S-matrix gives all orders in derivatives
with a fixed number of fields.  Both approaches yield crucial
information about the (2,1) string.

Let us begin with the sigma-model.  The (2,1) heterotic
sigma-model is most conveniently described in N=(2,0)
superfields.  These are of two types: bosonic chiral/antichiral 
superfields $X^i=x^i+\theta_R\chi^i+\ldots$ and
$X^{\bar i}=x^{\bar i}+\bar\theta_R\chi^{\bar i}+\ldots$
which are complex coordinates on the \tpt\ target space;
and fermionic chiral/antichiral superfields
$\Lambda^\alpha=\lam^\alpha+\theta_R F^\alpha+\ldots$ and
$\Lambda^{\bar \alpha}=
\lam^{\bar \alpha}+\bar\theta_R F^{\bar \alpha}+\ldots$
which give a fermionic description of the `internal space'
of the heterotic string.  The sigma-model action is
\beq
  S_{\sigma-\rm model}=\int d^2zd^2\theta_R[
	K_\mu(X,\bar X)\d X^\mu + h_{ab}(X,\bar X)
		\Lambda^a\Lambda^b]\ ,
\label{ssigma}
\eeq
where $\mu=i,\bar i$ and $a=\alpha,\bar\alpha$ collect together the 
complex indices.  Expanding in components, one finds a 
conventional heterotic sigma-model, with
\begin{eqnarray}
g_\ij&=&\d_i K_{\bar j}-\d_{\bar j}K_i\qquad\qquad g_{ij}=0\nonumber\\
b_\ij&=&\d_i K_{\bar j}+\d_{\bar j}K_i\qquad\qquad b_{ij}=0\nonumber\\
A_i&=&h^{\half}\d_i h^{-\half}\qquad\qquad 
	A_{\bar i}=(h^{-\half})\d_{\bar i} h^{\half}
\label{backgd}
\end{eqnarray}
\ie\ the fields are written in terms of prepotentials.
Note that the expression for $A_\mu$ is simply the Yang ansatz
for self-dual gauge fields.  
The left-moving N=1 supersymmetry is not manifest in this
formalism.  It will result in a relation between the background
metric and gauge fields;
henceforth, when we refer to the heterotic gauge field we will
mean the one for the `internal space', suppressing that part 
which is directly determined by the metric via the
left supersymmetry.
A fact that will be important later is
that conformal (2,0) sigma-models in the critical dimension $D=4$
automatically have (4,0) supersymmetry, given by a triplet of
complex structures $I_{\mu\nu}^{(r)}$, $r=1,2,3$, obeying
the algebra of imaginary quaternions:
\beq
I^{(r)}I^{(s)}=-\delta^{rs}+\eps^{rst}I^{(t)}\ .
\label{quats}
\eeq
Thus the target space is hyperK\"ahler (with torsion).

\subsection{\normalsize\bf $\beta$-functions}

Let us temporarily ignore the gauge field couplings 
in $S_{\sigma-\rm model}$,
equation \pref{ssigma}.  The one-loop beta-functions for the
purely gravitational sector are then
\beq
  R_{\mu\nu}(\Gamma)=\nabla_\mu\nabla_\nu\Phi\ ,
\label{betagrav}
\eeq
where $\Gamma^\mu_{\nu\lam}=\bigl\{ {\mu\atop\nu\lam} \bigr\}
-\hf H^{\mu}_{\nu\lam}$ is the connection with torsion.
This system of equations may be integrated \cite{hull}
to find (after a holomorphic coordinate transformation)
\begin{eqnarray}
e^{-2\Phi} \det[g]=1\nonumber\\
\Gamma_\mu\equiv\Gamma^\nu_{\lam\mu}I^\lam_\nu=0\ .
\label{fourzero}
\end{eqnarray}
The first equation determines the dilaton $\Phi$
algebraically in terms of the metric.  The second equation gives
the dynamics of the K\"ahler vector potential; it arises from the
variational principle
\beq
  S_{2+2}^{{\it grav}} = \int d^4x\sqrt{\det[g_{i\bar j}]}\ .
\label{sfourgrav}
\eeq
Now we add the heterotic gauge sector.  This comes from
a coupling to chiral fermions $\lam^a$
in the sigma-model, and results in the standard shifts
(again to one loop)
\begin{eqnarray}
\Gamma&\longrightarrow&\{{\rm Christoffel}\}-\hf(H+\omega_3^{{\rm YM}})
\nonumber\\
g_\ij&\longrightarrow& g_\ij+A_i^aA_{\bar j}^a\ .
\label{anomaly}
\end{eqnarray}
Note that there is no Lorentz Chern-Simons contribution to $\Gamma$,
because the N=1 left-moving supersymmetry pairs four $\lam^a$
with the four $x^\mu$ and cancels this part of the anomaly
which comes from the right-moving $\chi^\mu$.
Since the expression \pref{backgd} for $g_\ij$ in terms of
the K\"ahler vector potential $K_\mu$ admits an abelian gauge
invariance, we may write the expansion of the metric 
around flat space as 
$g_\ij=\eta_\ij+f_\ij$, with $f_\ij$ the ``field strength''
of the K\"ahler vector potential.

Taking the Yang-Mills gauge group $[U(1)]^8$ relevant to 
spacetime supersymmetric vacua of the (2,1) string
(\ie\ $A_i^a=i\d_i\phi^a$, $A_{\bar i}^a=-i\d_{\bar i}\phi^a$),
we achieve the result \cite{km2}
\beq
  S_{2+2} = \int d^4x\sqrt{\det[\eta_{i\bar j}
        +f_{i\bar j}+\d_i\phi^a\d_{\bar j}\phi^a]}\ .
\label{stwotwo}
\eeq
The action has the form of the Dirac-Born-Infeld
action for D-branes, but the determinant is {\it two-dimensional},
not four-dimensional!  This is exactly what one would
ask for the dimensional reduction to two dimensions that gives
target space strings, where for example one may set
$x^i=x^{\bar i}$.  However, one wonders what the dimensional
reduction to three dimensions has to do with the DBI action
of the two-brane and its three-dimensional determinant.
In \cite{kmo}, it was shown that to leading nontrivial order,
$S_{\tpt}$ reduced to D=2+1 agrees with the two-brane DBI action
after a coordinate transformation and a field redefinition.
It would be interesting to understand the precise relation.

\subsection{\normalsize\bf N=(2,1) string S-matrix}

So far, the expression \pref{stwotwo} for $S_{\tpt}$ is a
one-loop result.  For D-branes, the DBI action suffers higher-order
corrections in $\alpha'$.  However, analysis of the (2,1) string
S-matrix shows that $S_{\tpt}$ is {\it exact}.

The only nonvanishing S-matrix element of any N=2 string
is the three-point function, due to the incompatibility of
Regge behavior and the finite number of physical states.
The three-point function of the (2,1) string has the
schematic form

\setlength{\unitlength}{1mm}
\begin{picture}(80,40)
\put(10,20){\circle*{5}}
\put(10,20){\line(-5,3){5}}
\put(10,20){\line(5,3){5}}
\put(10,20){\line(0,-1){5}}
\put(20,15){\makebox(10,10){=}}
\put(40,20){\line(-5,3){5}}
\put(40,20){\line(5,3){5}}
\put(40,20){\line(0,-1){5}}
\put(55,15){\makebox(10,10){$\sim\quad k^3\xi^3$}}
\end{picture}

\noindent
\ie\ three fields $\xi$ and three powers of momenta.\footnote{The
(2,0) string also has a term $k^5\xi^3$, reflecting the
presence of a Lorentz Chern-Simons term.  This term invalidates
(for the (2,0) string) the power-counting argument which follows.}
The vanishing of the four-point function implies

\setlength{\unitlength}{1mm}
\begin{picture}(120,40)
\put(10,20){\circle*{5}}
\put(10,20){\line(-1,1){5}}
\put(10,20){\line(1,1){5}}
\put(10,20){\line(-1,-1){5}}
\put(10,20){\line(1,-1){5}}
\put(22,15){\makebox(15,10){$=\quad 0\quad =$}}
\put(50,20){\line(-1,1){5}}
\put(50,20){\line(-1,-1){5}}
\put(50,20){\line(1,0){5}}
\put(55,20){\line(1,1){5}}
\put(55,20){\line(1,-1){5}}
\put(70,15){\makebox(15,10){+ two more \hskip .5cm +}}
\put(100,20){\line(-1,1){5}}
\put(100,20){\line(-1,-1){5}}
\put(100,20){\line(1,1){5}}
\put(100,20){\line(1,-1){5}}
\end{picture}

\noindent
The nonlocal terms cancel among the $s$, $t$, and $u$
channel processes due to the special features of
the vertices and \tpt\ kinematics.  The leftover local term has the
scaling $k^3\cdot\frac1{k^2}\cdot k^3\sim k^4$, and must
be cancelled by a contact term of the form $k^4\xi^4$.
Proceeding in this way, one finds the $n$-point function has
the structure $k^n\xi^n$.  But this is exactly what one finds
in the expansion of $S_{\tpt}$!  A higher loop term in the beta-function
would imply terms in the S-matrix of the form $k^m\xi^n$ for
$m>n$, which is not seen.  We conclude that \pref{stwotwo}
is the {\it exact} bosonic target space effective action.
In order to verify these general arguments, the three- 
and four-point functions have been computed explicitly \cite{km2}
and indeed agree with $S_\tpt$.

\subsection{\normalsize\bf Fermion terms}

We have conducted a preliminary analysis of the fermion
(Ramond sector) couplings of the (2,1) string.  These are notoriously
difficult to analyze using sigma-model techniques (although
the recently developed Green-Schwarz-Berkovits formalism
may be helpful here \cite{berksieg,dbs}); therefore,
we have analyzed some of the three- and four-point S-matrix
elements.  Much of the result appears to be summarized
in the shifts
\begin{eqnarray}
  \d_i\phi^a\longrightarrow \d_i\phi^a+\bar\psi\gamma^a\d_i\psi
        \equiv\Pi_i^a\nonumber\\
  f_{i\bar j}\longrightarrow f_{i\bar j}+\bar\psi\gamma_{(i}
        \d_{\bar j)}\psi\equiv \Pi_{i\bar j}\ .
\label{gs}
\end{eqnarray}
This is in fact just what one expects for a super $p$-brane.
Recall the covariant super $p$-brane action
\beq
  S_{p-{\rm brane}}=\int d^{p+1}x\sqrt{\det[\Pi^a_\mu\Pi^a_\nu]}
	+\eps^{\mu_1\cdots\mu_{p+1}}
	[\Pi_{\mu_1}^{a_1}\cdots\Pi_{\mu_p}^{a_p}
	\bar\psi\gamma^{a_1\cdots a_p}\d_{\mu_{p+1}}\psi+\ldots]\ ,
\label{spbrane}
\eeq
where $\Pi_\mu^a=\d_\mu\phi^a-\bar\psi\gamma^a\d_\mu\psi$,
\ie\ $(\phi,\psi)$ are superspace coordinates, and $\Pi_\mu^a$ is
a tangent vector to superspace.  The first term in \pref{spbrane}
is the superspace volume element; combining \pref{stwotwo},
\pref{gs}, one sees essentially the same structure for the
M-brane, modulo the issue of 3D vs. 2D determinants
in the M-brane action discussed above.  The second term
in \pref{spbrane} is a superspace Wess-Zumino term, and is required
for $\kappa$-symmetry to hold.  It is not seen when the
fermions are set to zero, and hence would not have been 
found in the sigma-model approach above.  In the S-matrix approach,
one looks for parity-violating terms to establish this structure.

In the three-point function, one sees a term
\beq
  \int d^4x[f_{ij}(\bar\psi\gamma_{\bar k}\d_{\bar\ell}\psi)+h.c]
	\eps^{ij}\eps^{\bar k\bar\ell}
\label{swz}
\eeq
not accounted for by the Dirac-Born-Infeld term.  If for a moment
we relax the requirement of spacetime supersymmetry,
a nonabelian heterotic gauge sector of the (2,1) string
reveals a Wess-Zumino coupling of the transverse degrees of
freedom \cite{ov}
\beq
  \int_{\d\MM_5=\MM_4}d^5x\; I\wedge tr[(h^{-1}dh)^3]\ ,
\label{wz}
\eeq
where $I$ is the K\"ahler form.  Thus one sees 
that Wess-Zumino terms are generic, and 
that there is the possibility
of the right sort of Wess-Zumino terms appearing in the M-brane
action so that it admits a $\kappa$-symmetry.  
It is important to complete
the determination of the WZ terms -- they are the expectation values
of superspace antisymmetric tensor fields, and their evaluation
will go a long way towards determining what sort of spacetime fields
couple to the M-brane.

Finally, it is curious to note that a remarkable transmutation 
has taken place in $S_\tpt$ \pref{stwotwo}.  One begins with
gauge fields $g_{\mu\nu}$, $b_{\mu\nu}$, $A_\mu^a$; imposing
the self-duality constraints of global (4,0) supersymmetry,
the effective action is written in terms of the potentials
$K_\mu$, $\phi^a$ in a gauge fixed, yet still remarkably
geometrical form.  Similarly, the fermions start life as
a gravitino field, and end up as superspace coordinates!

\section{\large\bf Symmetries, self-duality, and integrability}

In this section we will discuss the properties, both
known and expected, of $S_\tpt$.  We begin with an expected
symmetry, $\kappa$-symmetry; then consider a known one,
self-duality, and its associated symmetries and conservation laws.

\subsection{\normalsize\bf Supersymmetry and $\kappa$-symmetry}

The (2,1) string vertex algebra has manifest {\it linear}
target space supersymmetry \cite{fms}, which is the dimensional
reduction from ten or twelve dimensions down to the four-dimensional
M-brane world volume:
\begin{eqnarray}
  \delta K_\mu&\sim&\bar\eps\gamma_\mu\psi\nonumber\\
  \delta \phi^a&\sim&\bar\eps\gamma^a\psi\nonumber\\
  \delta \psi&\sim&(f_{\mu\nu}\gamma^{\mu\nu}+\d_\mu\phi^a
	\gamma^\mu\gamma^a)\eps\ .
\label{neqfour}
\end{eqnarray}
There are higher-order corrections to these transformation laws
(\cf\ \cite{tseytlin} for a similar situation) due to the
nonlinearity of $S_\tpt$.  These transformations are to be
compared with the $\kappa$-symmetry and (nonlinearly realized
\cite{hupo}) supersymmetry of super $p$-branes (\cf\ \cite{BSTrev})
\begin{eqnarray}
  \delta\psi &=&\eta+(1+\gamma_{p+2})\kappa\nonumber\\
  \delta\phi^a&=&\bar\eps\gamma^a\eta +
	\bar\psi\gamma^a(1+\gamma_{p+2})\kappa\ .
\label{gssusy}
\end{eqnarray}
In simpler situations, such as the Green-Schwarz string (\cf\ \cite{gsw})
and the supermembrane in D=4 \cite{achucarro}, it is known that
linear supersymmetry like \pref{neqfour} appears as a specific
combination of $\eta$- and $\kappa$-supersymmetries when working
in a physical gauge.  This gives a strong indication that such
a symmetry underlies a covariant formulation of the M-brane.
We also see that the M-brane action reduced to 2$+$1 dimensions
and the supermembrane action have the same spectrum
of spontaneously broken translations and supersymmetries (in fact,
one can {\it define} the $p$-brane action as the nonlinear lagrangian
for partially broken global supersymmetry \cite{hupo}),
and therefore must be related by some combination of field
redefinitions and field-dependent coordinate transformations
(although it is not ruled out that something more indirect,
such as a Backl\"und transformation, is involved).

\subsection{\normalsize\bf 
Self-duality, integrability, and conservation laws}

As mentioned above, the gauged U(1) of the right-moving N=2
local supersymmetry in the (2,1) string sits inside a
global N=4 hyperK\"ahler structure.  The integrable nature of
the theory is elucidated by considering the family
$\Pone = SU(2)/U(1)$
of complex structures,\footnote{In 4$+$0 signature; in \tpt\
signature one has the disc $SU(1,1)/U(1)$.} 
thus passing to the twistor space $\TT$,
the bundle $\Pone\rightarrow\MM_4$ whose coordinates
$u\in\Pone$, $z_u^i=z^i+u\eps^{ij}z^{\bar j}\in\MM_4$
define a natural complex structure.  For instance, in 
self-dual Yang-Mills (SDYM), given a transition function
$g\in G_{\IC}$ on the equator of $\Pone$, one may factorize
$g=f_+^{-1}f_-$, where $f_+(u,z^i_u)$ is holomorphic in the 
upper hemisphere and $f_-(u,z^i_u)$ is holomorphic in the
lower hemisphere.  Then one reconstructs the gauge field on $\MM_4$
from this holomorphic bundle over $\TT$ via
$\bar A=-(\bar\d h)h^{-1}$, where $h=f_+(u=0,z^i)$.

The twistor space relevant to (4,0) sigma-model geometry was
recently constructed by Howe and Papadopolous \cite{hopa}.
However, the details of the twistor transform remain to be worked out.
In principle, the complete classical physics of waves on the
M-brane is under control, and it is an interesting question
what the classical solutions look like.

Another interesting feature of the twistor construction for SDYM
is that it naturally involves the complexification of the target
space (one works over $G_{\IC}$, with a gauge symmetry that
reduces the dynamics in the end to a real slice).  We have also
seen a complexification of the world sheet in $S_\tpt$
(a different sort of `complexification' was investigated in
\cite{bvm}).  These properties are intriguing given that
Witten \cite{witten} invoked complexification of both worldsheet
and spacetime in order to explain certain features of string
behavior at high temperature and high energy.  It may be
that (2,1) strings provide a window into this regime.

Gursey and collaborators \cite{gursey} have drawn analogies
between 2D conformal symmetry as complex analyticity,
and 4D self-duality as a kind of `quaternionic analyticity'.
Indeed, many of the formulae of the ADHM construction of solutions
to SDYM are most conveniently written in terms of quaternions.
We will exploit this connection in section six when we speculate
on the ultimate structure of M-theory.

Losev \etal\ \cite{lmnsone,lmnstwo} have pointed out the strong
parallel between the twistor construction and the holomorphic
factorization property of rational conformal field theory.
Indeed, the SDYM equations have a symmetry under
$h(x,\bar x)\rightarrow f_L(\bar x)h(x,\bar x)f_R(x)$
related to a 4D generalization of the Polyakov-Wiegmann identity
\beq
  \Gamma[gh]=\Gamma[g]+\Gamma[h]
	-\coeff{i}{2\pi}\int_{\MM_4} I\wedge tr[g^{-1}\d g\cdot
		\bar\d h\;h^{-1}]\ ,
\label{pw}
\eeq
leading to corresponding current algebra and Ward identities.
The gravitational action $\int d^4x\sqrt{\det[g_\ij]}$ has a
similar residual symmetry under
$g_\ij\rightarrow\d_i f^k(x)g_{k\bar\ell}(f,\bar f)\d_{\bar j}
f^{\bar\ell}(\bar x)$ with $|\det\d_i f^j|=1$, 
\ie\ area-preserving holomorphic diffeomorphisms; this is an
analogue of conformal invariance in the present situation.
We again see that M-brane geometry is an extremely natural
generalization of string worldsheet geometry.
These symmetries generate infinite towers of conservation laws
which will help constrain the quantum theory.

If self-dual Yang-Mills generalizes rational conformal field theory
to a 4D context, what about other string backgrounds?
We now know of vast classes of nontrivial fixed points 
in 4D field theory, \eg\ \cite{arg}.
One ought to be able to suitably twist these to construct 
self-dual theories; one might obtain the analogue of
Landau-Ginsburg models \cite{mart} in this way.

\section{\large\bf Quantization}

While we have not thought deeply about the quantization of M-branes,
it seems evident that a quantum theory exists, since
the dynamics is essentially trivial -- the theory is integrable,
the S-matrix more or less vanishes, and of course N=(2,1)
strings manifestly generate a perturbative quantization.
However, the quantum theory of N=(2,1) strings is not necessarily
a field theory.  The first indication of this comes from the
power-counting of N=2 string loops, which is two-dimensional
(instead of four-dimensional) due to an extra factor of the 
Schwinger parameter $\tau$ coming from the normalization of the
N=2 string U(1) modulus.  Another indication is that momenta in
the internal directions (if we bosonize the $\lambda^a$
to make the $E_8$ torus) flows off-shell in loops;
in fact, once one of the \tpt\ target coordinates is
compactified on a circle, a rich spectrum of {\it physical}
states arises \cite{kmu}, whose interpretation
is at the moment obscure.  These are the analogues for the (2,1)
string of the perturbative BPS states of the usual (1,0)
heterotic string; when both momenta and winding are present
on the circle, level matching permits the right-moving
N=2 sector to remain in its ground state while the left-moving
N=1 sector has essentially arbitrary excitation.  Also,
in the case where only winding is present on the circle,
one gets a {\it second copy} of the target 
space M-brane \cite{km,bvm}.

There are some intriguing analogies between the (2,1) string
on the one hand, and matrix models of noncritical strings
\cite{2dgrav} on the other (see figure 2).
Two-dimensional noncritical string theory is almost a field theory
of the center-of-mass (``tachyon'') degree of freedom, but
not quite.  For instance, when compactified on a circle,
there are both `momentum tachyons' and `winding tachyons'; the
partition function has an $R\rightarrow1/R$ symmetry.
As we have just seen, the (2,1) string also has this sort
of doubling.  

\vskip .2cm
\begin{tabular}{|c|c|}
\hline
 Matrix models/noncritical strings & N=2 strings \\
\hline
tachyon field theory & $S_{\tpt}[\Phi]$ \\
cubic collective field theory & $W[J]=S_{\tpt}-J\cdot \Phi$ \\
matrix model Fermi fluid & ?? \\
\hline
\end{tabular}
\vskip .5cm
\centerline{Figure 2. \it Analogies between noncritical and
N=2 strings.}
\vskip .3cm

\noindent
There is a nonpolynomial field theory for the
noncritical string tachyon field (the analogue of $S_\tpt$).
This field theory has a representation in terms of a cubic collective
field theory, which is obtained via an integral transform of
the tachyon field and encodes much of the symmetry and
integrability.  The analogue for the (2,1) string might be the
Legendre transform to sources, which would be cubic on-shell
since all of the $n$-point amplitudes vanish for $n\ge 4$;
or perhaps the twistor transform, which linearizes the theory.
For the noncritical string, 
neither of these presentations 
is appropriate to the full quantum theory, however, which is
best written in terms of the fermi fluid of the matrix model.
Thus one wonders whether there might be such a simpler
presentation appropriate to the quantization of (2,1) strings.
Since the classical theory is solved on twistor space,
presumably one should look for some sort of matrix dynamics whose
semiclassical limit is naturally expressed in terms of twistors,
and whose collective excitations are (2,1) strings.

Of course, an alternate interpretation is that N=(2,1) string
theory has some overlap with M-theory but is not fundamental
to it.  One might regard the above observations as evidence
to that effect: The doubling of the spectrum, the BPS-like tower,
the fact that the internal space is restricted to be the
$E_8$ torus, etc.  The constructions of \cite{km,kmo} realize
a particular special class of states in particular M-theory
backgrounds, but certainly do not generate all of the theory.
Perhaps N=(2,1) strings know about M-theory because they
know about (a) spacetime supersymmetry, and (b) self-dual
worldvolumes.  This would lead us to ignore all this excess
baggage carried by (2,1) strings, and to attempt to quantize
directly the \tpt\ field theory $S_{\tpt}$.
I regard the developments of \cite{lmnsone,lmnstwo}
as promising steps in this direction.

\section{\large\bf Twelve dimensions (\ie\ \tentwo)}

The spacetime supersymmetry algebra of the (2,1) string
is generated by world sheet charges \cite{km,kmo}
which schematically take the form
\beq
Q_\alpha=\oint dz\;\Sigma_{\rm ghost}S_\alpha\ ,
\label{susyq}
\eeq
and obey the algebra
\beq
\{Q_\alpha,Q_\beta\}=\oint\widetilde\Sigma_{gh}\lam^a\lam^b
	(\gamma^{ab})_{\alpha\beta}\ .
\label{tentwoalg}
\eeq
Here $S_\alpha$ is an O(10,2) Majorana-Weyl spinor, the spin
field of the heterotic fermions $\lam^a$; thus $\lam^a\lam^b$
is the O(10,2) generator on this part of the theory.
However, the theory is not O(10,2) invariant; the BRS
constraints impose the null reduction $\slash v Q=0$
for some null vector $v$.  In the supersymmetry algebra
\pref{tentwoalg} this forces one of the polarizations $a,b$
on the RHS to point along the null vector; the remaining
operator is a picture change of the momentum.  Thus the physical
states obey
\beq
  \{Q_\alpha,Q_\beta\}=P^a v^b(\gamma^{ab})_{\alpha\beta}\ .
\label{phystentwo}
\eeq
The RHS realizes a particular state in the natural \tentwo\
supersymmetry algebra \cite{vhvp}
\beq
  \{Q_\alpha,Q_\beta\}=M^{ab}(\gamma^{ab})_{\alpha\beta}
	+ Z_+^{a_1\cdots a_6}(\gamma^{a_1\cdots a_6})_{\alpha\beta}\ ,
\label{tentwosusy}
\eeq
where the bosonic charges are evaluated as $M^{ab}=P^{[a}v^{b]}$,
$Z_+^{a_1\cdots a_6}=0$.  This algebra is to be contrasted with
10+1 supersymmetry
\beq
  \{Q_\alpha,Q_\beta\}=P^a(\gamma^{a})_{\alpha\beta}
	+ M^{ab}(\gamma^{ab})_{\alpha\beta}
        + Z^{a_1\cdots a_5}(\gamma^{a_1\cdots a_5})_{\alpha\beta}\ .
\label{tenonesusy}
\eeq
Comparing this algebra with \pref{tentwosusy}, one sees
that the 11 $P^a$ and 55 $M^{ab}$ of 10+1 dimensions have been
``unified'' into the 66 $M^{ab}$ of \tentwo\ dimensions
(the 11D five-form and 12D self-dual six-form both have 462
components).  Since $P^a$ couples to the metric $e_{\mu a}$ while
$M^{ab}$ couples to the antisymmetric tensor gauge field
$A_{\mu ab}$, an
unbroken realization of this symmetry would unify all electric
charges.  Such a unification might explain why scalars coming
from the antisymmetric tensor and the metric form multiplets
of a larger symmetry (U-duality) in lower dimensions
(including exceptional groups; see below).
Actually, the unbroken algebra \pref{tentwosusy} would imply
a complete `$p$-brane democracy' \cite{townsend},
since the algebra \pref{tentwosusy} is OSp(1$|$32);
the commutator of the $M^{ab}$ closes on the $Z_+^{a_1\cdots a_6}$.
Clearly such a structure is incompatible with flat spacetime,
and is not seen in the N=2 string.  Yet it is tempting to ask
whether O(10,2) is related to the conformal group, spontaneously
broken to O(9,1).  Then the Planck scale would be dynamically
generated (and perhaps related to the N=(2,1) string tension?),
and the high energy behavior governed by scale-invariant
physics and the unbroken realization of \pref{tentwosusy}.
Indeed, quantum cohomology provides a clue that the distinction
between different degree forms disappears at short distances
(\eg\ K3 moduli space in string theory is O(20,4), not
just the O(19,3) that acts on $H^2(\Z)$;
the extra generators mix zero-, two-, and four-forms
for small volume K3's).

\section{\large\bf 12+4 and beyond}

The question arises whether there is any limit to adding hidden
dimensions to a theory.  I think the N=(2,1) string reveals
sufficient additional structure to provide an answer to this
question.\footnote{The ideas presented here
were initiated in discussions with P.G.O. Freund.}
The quest for unification has seen a progression
from particle world lines ($\IR$ analyticity) of quantum
field theory, to string world sheets ($\IC$ analyticity)
in order to incorporate gravity, and now M-brane
worldvolumes ($\IH$ analyticity) in order to realize duality.
Clearly there is room for one further step, to octonionic
($\IO$) analyticity.  Just as supersymmetry picks out transverse
dimensions $\nu=1,2,4,8$ -- corresponding to the division
algebras $\IK_\nu=\IR,\IC,\IH,\IO$ --
we are now exploring symmetries of longitudinal directions.
In fact, there is a `magic square' of orthogonal algebras
\cite{sudbery} (based on $H_2(\IK_\nu)=2\times2$ Hermitean
matrices over $\IK_\nu$)

\vskip .2cm
\begin{tabular}{c||l||c|c|c|c||rl}
\multicolumn{8}{c}{transverse $\longrightarrow$}\\
\cline{2-6}
& & $\IR$ & $\IC$ & $\IH$ & $\IO$ & & \\ \cline{2-6}
longitudinal & $\widetilde \IR$ & O(2) & O(3) & O(5) & O(9) &
	{\bf sa}(2,$\IK_\nu$) & rotation  \\ \cline{2-6}
$\downarrow$ & $\widetilde {\IC}$ & O(2,1) & O(3,1) & O(5,1) & O(9,1) &
	{\bf sl}(2,$\IK_\nu$) & Lorentz  \\ \cline{2-6}
& $\widetilde \IH$ & O(3,2) & O(4,2) & O(6,2) & {\bf O(10,2)} &
	{\bf sp}(4,$\IK_\nu$) & conformal  \\ \cline{2-6}
& $\widetilde {\IO}$ & O(5,4) & O(6,4) & O(8,4) & {\bf O(12,4)} &
	 ?\ \ \ \   & \ \ \ \ ?  \\ \cline{2-6}
\end{tabular}
\vskip .2cm

\noindent
These groups are constructed in the same manner as the usual
Tits-Freudenthal magic square, which uniformly constructs the 
exceptional groups using $H_3(\IK_\nu)=3\times3$ Hermitean
matrices over $\IK_\nu$:

\vskip .2cm
\hskip 2cm\vbox{
\begin{tabular}{||l||c|c|c|c||}
\hline
& $\IR$ & $\IC$ & $\IH$ & $\IO$   \\ \hline \hline
$\widetilde \IR$ & SO(3) & SU(3) & Sq(3) & $\rm F_4$ 
	\\ \hline
$\widetilde {\IC}$ & SL(3,$\IR$) & SL(3,$\IC$) & SL(3,$\IH$) &
		$\rm E_6$ 
	\\ \hline
$\widetilde \IH$ & Sp(6,$\IR$) & SU(3,3) & Sp(6,$\IH$) &
		$\rm E_7$ 
	\\ \hline
$\widetilde {\IO}$ & $\rm F_4$ & $\rm E_6$ & $\rm E_7$ & $\rm E_8$ 
	\\ \hline
\end{tabular}
}
\vskip .2cm

\noindent
Given the intimate association of the exceptional groups to
octonions, it is quite possible that octonionic worldvolume structures
could provide an explanation for the appearance of exceptional
groups in U-duality.

There are a few indications of 12+4 structure in the (2,1)
string.  First, the construction of the (2,1) string worldsheet
as the target space of the (2,1) string turns out to have
N=8 global right-moving supersymmetry, not just N=4.
Second, Moore \cite{moore} has suggested that maximal
symmetry in string theory arises when treating left- and 
right-movers independently, via Narain compactification
of all dimensions.  Applying this reasoning to the
(2,1) string leads to the Narain group O(12,4) from
\tentwo\ left-movers, and \tpt\ right-movers.  
This symmetry becomes apparent in the regime where all radii
are of order the string scale of the underlying (2,1) string,
lending further support to the idea of symmetry
restoration at short distances.  Finally, one will need
additional degrees of freedom to describe the M-brane 
in a reparametrization invariant way.  The reparametrization
invariant 11D supermembrane action utilizes 10+1 scalars
(8 transverse and 2+1 longitudinal).  Dualizing a transverse
scalar on the worldvolume yields 9+1 scalars (7 transverse
and 2+1 longitudinal) and the 2+1 components of a vector
\cite{townschmid}.
This parametrization is the one generated by the N=(2,1) string,
albeit in static gauge where the 2+1 longitudinal scalars
have been eliminated by gauge-fixing.  A covariant description
of the M-brane before reparametrization gauge fixing and null
reduction would require \tentwo\ scalars
(8 transverse and \tpt\ longitudinal) as well as the \tpt\ 
components of a vector. 
If present, octonionic symmetry would have to relate the
vector and the scalars; however this is just what is called for
if one is to realize the O(10,1) symmetry of 11D supergravity.

Additional spacelike coordinates may be required to
match expected hidden structure of the IIB theory
\cite{hull2,vafa}, which seems to require 11+1 signature.
Recall that our construction realizes {\it both} the IIB
string and the IIA two-brane within the moduli space of
(2,1) strings \cite{km,kmo}.  How can we get the IIB string,
with its supercharges both $\underline {16}$'s of the same
chirality in O(9,1), when the supersymmetry 
charges \pref{susyq} form a single $\underline {32}$
of O(10,2)?  The answer lies in the fact that we are in
static gauge, where only half of the supersymmetry is
linearly realized; it turns out that,
for both the supermembrane and the IIB string, the unbroken
supercharges in static gauge
are compatible with the null reduction of
an O(10,2) spinor.   However, a covariant formulation
would have to accomodate both IIA and IIB superalgebras;
O(12,4) spinors have plenty of room 
(the fermionic partners of the 12+4 bosons would fill a 
$\underline {128}$ of O(12,4)).
Related to this is the fact that the eleventh dimension of
IIA supergravity comes in our construction from the
field space of the vector; the ten spatial dimensions 
of 11D supergravity are {\it not}
the same as the ten spatial dimensions of \tentwo, one of which
is always eliminated in the null reduction.

\section{\large\bf Conclusions}

The construction of M-branes from N=(2,1) heterotic strings
encodes key aspects of duality already in the classical
theory, indicating that we are approaching the underlying
degrees of freedom of the theory.  In particular, more
of the symmetry is manifest.  The worldvolume structure
is a `complexification' of string worldsheets whereby
4D self-duality ($\IH$ analyticity) supplants 2D conformal
invariance ($\IC$ analyticity) as the world-volume symmetry
principle.  There are a few indications that an extension
to octonionic analytic structures might be possible.
I expect that M-branes will prove to be as useful a probe of
structure in M-theory as D-branes and $p$-branes have.
We do not know how to quantize the latter objects 
either, yet they have been immensely helpful in our 
understanding of geometry, duality, and nonperturbative physics.
It is premature to say whether and how M-branes are the
`fundamental' objects of M-theory.  
Even so, their many magical properties warrant further 
investigation.

\vskip 1cm
\noindent {\bf Acknowledgements:} 
It goes without saying that the ideas presented here
are the product of a very fruitful collaboration with 
D. Kutasov, and also M. O'Loughlin.  Discussions with
P.G.O. Freund resulted in the speculations of section six.
I would also like to thank C. Hull for explanations
of OSp(1$|$32), and G. Moore for explanations of
his work \cite{lmnsone,lmnstwo}.
Thanks to the Aspen Center for Physics for hospitality during
the preparation of this manuscript.
This work is supported in part by funds provided by the DOE under
grant No. DE-FG02-90ER-40560.

\vskip 1cm
\noindent {\bf Note added:}
At the conference, C. Hull informed me that he had obtained
the gravitational part of $S_\tpt$; I. Bars \cite{bars}
presented a framework unifying both IIA and IIB
superalgebras, starting from 11+2 dimensions.  He also
found the realization of the \tentwo\ superalgebra in
equation \pref{phystentwo}.  L. Susskind presented work
relating M-theory to SU(N) matrix mechanics along the lines
of \cite{dhn,dln}, giving arguments that the standard
11D supermembrane is obtained as a collective excitation
in a particular semiclassical limit.
G. Chapline pointed out his work relating fermi fluids
and self-dual gravity \cite{chapline}, and other work
related to \cite{lmnstwo}.
After the conference, work of Jevicki has appeared
\cite{jevicki} relating self-dual gravity to large N
matrix mechanics; and Nishino and Sezgin \cite{ns}
also have investigated the realization \pref{phystentwo}
of \tentwo\ supersymmetry.

\end{document}